\documentclass[epj]{svjour}
\usepackage{graphicx}
\usepackage{amssymb,amsbsy,amsmath}

\newcommand{\op}[1]{%
    \fontdimen12\textfont3=2pt\fontdimen12\scriptfont3=1.4pt%
    \!\null\mathop{\vphantom{#1}\smash{#1}}\limits_{\sim}\null\!}
\newcommand{\xref}[1]{\protect\ref{#1}}
\newcommand{\figref}[1]{Fig.~\protect\ref{#1}}
\newcommand{\fmref}[1]{(\protect\ref{#1})}
\def\bra#1{\langle \, {#1} \, | \,}
\def\ket#1{\, | \, {#1} \, \rangle}
\newcommand{\braket}[2]{\langle \, {#1} \, | \, {#2} \, \rangle}
\journalname{Eur. Phys. J. B}
\begin{document}
\title{Properties of highly frustrated magnetic molecules
  studied by the finite-temperature Lanczos method}
\titlerunning{Magnetic molecules studied by FTLM}
\author{J\"urgen Schnack%
 \and Oliver Wendland%
}                     
\offprints{J\"urgen Schnack}          
\institute{Department of Physics, Bielefeld University, P.O. box 100131, D-33501 Bielefeld, Germany}
\date{Received: date / Revised version: date}
%
\abstract{ The very interesting magnetic properties of
frustrated magnetic molecules are often hardly accessible due to
the prohibitive size of the related Hilbert spaces. The
finite-temperature Lanczos method is able to treat spin systems
for Hilbert space sizes up to $10^9$. Here we first demonstrate
for exactly solvable systems that the method is indeed
accurate. Then we discuss the thermal properties of one of the
biggest magnetic molecules synthesized to date, the
icosidodecahedron with antiferromagnetically coupled spins of
$s=1/2$. We show how genuine quantum features
such as the magnetization plateau behave as a function of
temperature.
\PACS{
{75.10.Jm}{Quantized spin models}   \and
{75.40.Mg}{Numerical simulation studies}   \and
{75.50.Xx}{Molecular magnets}
     } 
} 
\maketitle
%

\section{Introduction}
\label{sec-1}

The magnetism of antiferromagnetically coupled and geometrically
frustrated magnetic molecules
\cite{BGG:JCSDT97,MSS:ACIE99,SHS:PRL02,MTS:AC05,SNS:PRL05,TEM:CEJ06,SRC:PRB06,TMB:ACIE07,PLK:CC07}
is a fascinating subject due to the richness of phenomena that
are observed \cite{Ram:ARMS94,Gre:JMC01} as well as due to the
similarities that can be drawn towards extended spin systems
such as the two-dimensional kagom\'{e} lattice
\cite{SHS:PRL02,Gre:JMC01,Diep94,NKH:EPL04,Zhi:PRL02,Atw:NM02}.
But although mo\-lecules constitute finite-size spin systems,
the investigation of their magnetic properties for instance in
the Heisenberg model -- as function of both temperature and
magnetic field -- is largely restricted if not impossible due to
the enormous size of the underlying Hilbert spaces. Quantum
Monte Carlo (QMC) calculations are of no help in this case since
they suffer from the so-called negative-sign problem
\cite{SaK:PRB91,San:PRB99,EnL:PRB06}. Density Matrix
Renormalization Group (DMRG) techniques provide another very
powerful approximation mainly for one-dimensional spin systems
such as chains \cite{Whi:PRB93,Sch:RMP05}. The method delivers
the relative ground states for orthogonal subspaces. Extensions
to include the approximate evaluation of excitations have been
developed recently \cite{Jec:PRB02}. Nevertheless, the whole
method still works best for one-dimensional systems;
applications to magnetic molecules are rare \cite{ExS:PRB03}.

A method, which can treat medium size spin systems irrespective
of their geometric structure, is the Lanczos method
\cite{Lan:JRNBS50}. This method yields eigenstates with extremal
eigenvalues in orthogonal subspaces with high accuracy and is
thus able to deliver a magnetization curve at $T=0$. An
extension towards $T>0$ is the finite-temperature Lanczos method
(FTLM) \cite{PhysRevB.49.5065}. Although it was applied to
several Heisenberg or Hubbard model systems, see e.g.
\cite{PhysRevB.49.5065,JaP:AP00,PhysRevB.67.161103,ZST:PRB06,PhysRevB.76.125113,PhysRevB.79.115141},
one must say, that this method is not yet very common.

In this article we investigate whether the finite-tem\-perature
Lanczos method (FTLM) is applicable for the Heisenberg model
describing magnetic molecules. To this end its accuracy is first
compared for exactly solvable cases. Thanks to recent advances
in the application of group theoretical methods, the energy
spectra of spin systems of unprecedented size can be evaluated
numerically exactly \cite{ScS:IRPC10}. Thus,
antiferromagnetically coupled spin systems with the geometric
structure of the cuboctahedron and the icosahedron with $s=3/2$
will serve as test cases; the Hilbert space dimension is
16,777,216 for both \cite{ScS:IRPC10,ScS:PRB09,ScS:P09}.

Finally, the finite temperature behavior of an
antiferromagnetically coupled spin system with the geometric
structure of the icosidodecahedron, that is closely related to
the kagom\'{e} lattice, will be examined for $s=1/2$ for the
first time. Although its total Hilbert space dimension is
1,073,741,824, the finite-temperature Lanczos method is able to
deliver the magnetization and the heat capacity as function of
both temperature and applied magnetic field.

The article is organized as follows. In Section~\xref{sec-2}
basics of the finite-temperature Lanczos method are
repeated. Section~\xref{sec-3} is devoted to the discussion of
the accuracy of the method, and in Section~\xref{sec-4} the
method is applied to the icosidodecahedron. The article closes
with a summary.

\section{Reminder of the finite-temperature Lanczos method}
\label{sec-2}

For the evaluation of thermodynamic properties in the canonical
ensemble the exact partition function $Z$ depending on
temperature $T$ and magnetic field $B$ is given by 
\begin{eqnarray}
\label{E-1-1}
Z(T,B)
&=&
\sum_{\nu}\;
\bra{\nu} e^{-\beta \op{H}} \ket{\nu}
\ .
\end{eqnarray}
Here $\{\ket{\nu}\}$ denotes an orthonormal basis of the
respective Hilbert space. Following the ideas of
Refs.~\cite{PhysRevB.49.5065,JaP:AP00} the unknown matrix
elements are approximated as
\begin{eqnarray}
\label{E-1-2}
\bra{\nu} e^{-\beta \op{H}} \ket{\nu}
&\approx&
\sum_{n=1}^{N_L}\;
\braket{\nu}{n(\nu)} e^{-\beta \epsilon_n^{(\nu)}} \braket{n(\nu)}{\nu}
\ ,
\end{eqnarray}
where $\ket{n(\nu)}$ is the $n$-th Lanczos eigenvector starting
from $\ket{\nu}$ as the initial vector of a Lanczos
iteration. $\epsilon_n^{(\nu)}$ denotes the associated $n$-th
Lanczos energy eigenvalue. The number of Lanczos steps is chosen
as $N_L$. In addition, the complete and thus very large sum over all states
$\ket{\nu}$ is replaced by a summation over a subset of $R$
random vectors. Altogether this yields for the partition function
\begin{eqnarray}
\label{E-1-3}
Z(T,B)
&\approx&
\frac{\text{dim}({\mathcal H})}{R}
\sum_{\nu=1}^R\;
\sum_{n=1}^{N_L}\;
e^{-\beta \epsilon_n^{(\nu)}} |\braket{n(\nu)}{\nu}|^2
\ .
\end{eqnarray}
Although this already sketches the general idea, it will always
improve the accuracy if symmetries are taken into
account as in the following formulation
\begin{eqnarray}
\label{E-1-4}
Z(T,B)
&\approx&
\sum_{\Gamma}\;
\frac{\text{dim}({\mathcal H}(\Gamma))}{R_{\Gamma}}
\sum_{\nu=1}^{R_{\Gamma}}\;
\sum_{n=1}^{N_L}\;
\nonumber \\
&&\times
e^{-\beta \epsilon_n^{(\nu,\Gamma)}} |\braket{n(\nu, \Gamma)}{\nu, \Gamma}|^2
\ .
\end{eqnarray}
Here $\Gamma$ labels the irreducible representations of the
employed symmetry group. The full Hilbert space is decomposed
into mutually orthogonal subspaces ${\mathcal H}(\Gamma)$.

An observable would then be calculated as
\begin{eqnarray}
\label{E-1-5}
O(T,B)
&\approx&
\frac{1}{Z(T,B)}
\sum_{\Gamma}\;
\frac{\text{dim}({\mathcal H}(\Gamma))}{R_{\Gamma}}
\sum_{\nu=1}^{R_{\Gamma}}\;
\sum_{n=1}^{N_L}\;
e^{-\beta \epsilon_n^{(\nu,\Gamma)}}
\nonumber \\
&&\times
\bra{n(\nu, \Gamma)}\op{O}\ket{\nu, \Gamma}
\braket{\nu, \Gamma}{n(\nu, \Gamma)}
\ .
\end{eqnarray}
It was noted in Ref.~\cite{PhysRevB.67.161103} that this
approximation of the observable $O(T,B)$ may contain large
statistical fluctuations at low temperatures due to the
randomness of the set of states $\{\ket{\nu, \Gamma}\}$. It was
shown that this can largely be cured by assuming a symmetrized
version of Eq.~\fmref{E-1-5}. For our investigations this is
irrelevant.

The very positive experience is that even for large problems the
number of random starting vectors as well as the number of
Lanczos steps can be chosen rather small, e.g. $R\approx 20,
N_L\approx 100$. The later sections will provide further
evidence for this statement.

It is foreseeable that the method does not work optimally in
very small subspaces or subspaces with large degeneracies of
energy levels especially if the symmetry is not broken up into
irreducible representations $\Gamma$. The underlying reason is
given by the properties of the Lanczos method itself that fails
to dissolve such degeneracies. The other case of small subspaces
can be solved by including their energy eigenvalues and
eigenstates exactly.

Another technical issue is given by the fact that the chosen
random vectors $\ket{\nu, \Gamma}$ should be mutually
orthogonal. Although one could orthonormalize the respective
vectors, this is for practical purposes not really
necessary. The reason is, that two vectors with random
components are practically always orthogonal, because their
scalar product is a sum over fluctuating terms that nearly
vanishes especially in very large Hilbert spaces.

Since Lanczos iterations consist of matrix vector
multiplications they can be parallelized by \verb§openMP§
directives. In our programs this is further accelerated by an
analytical state coding and an evaluation of matrix elements of
the Heisenberg Hamiltonian ``on the fly" \cite{SHS:JCP07}.

\section{Cuboctahedron and Icosahedron}
\label{sec-3}

\begin{figure}[ht!]
\centering
\includegraphics*[clip,width=40mm]{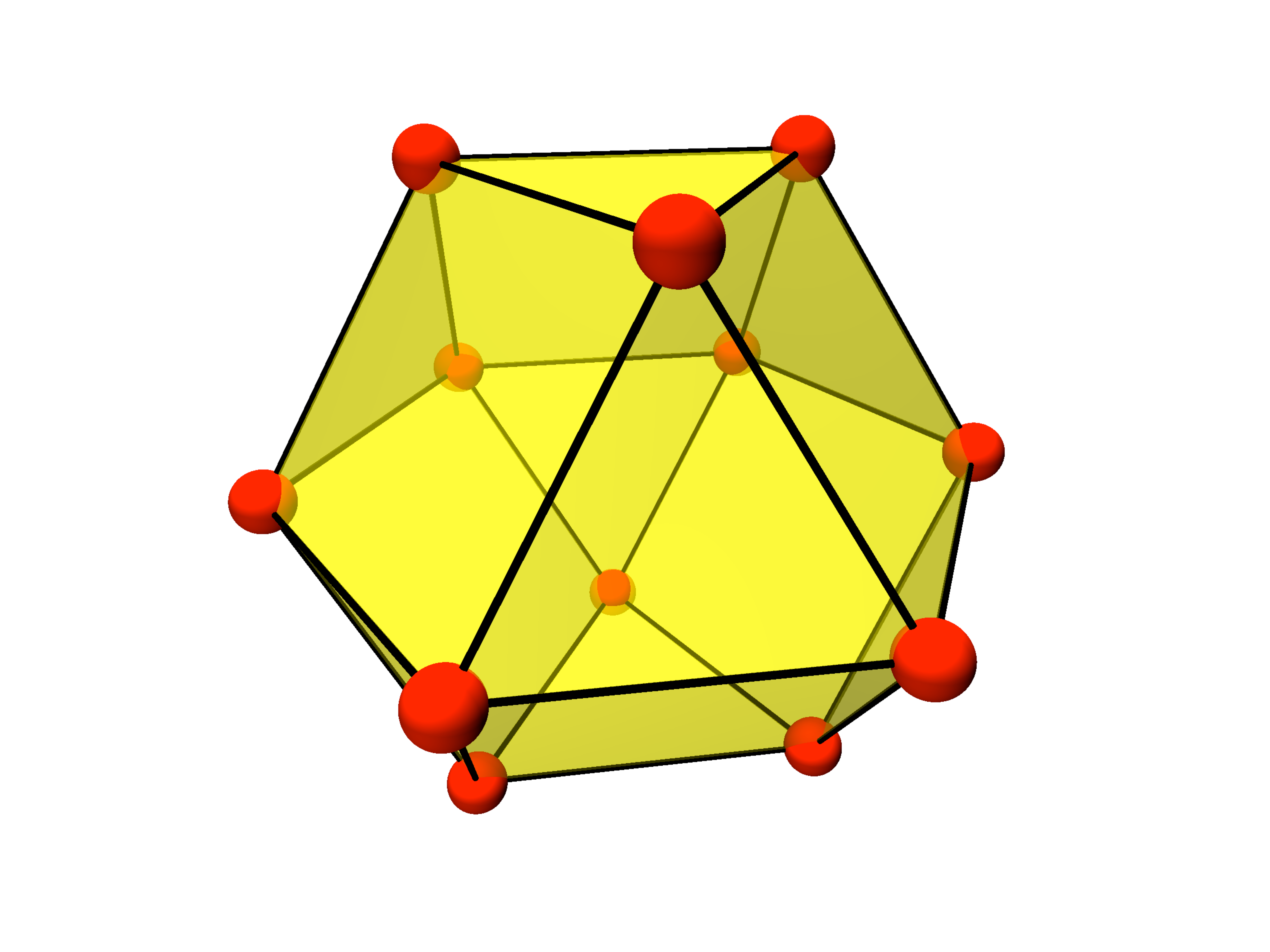}
\includegraphics*[clip,width=40mm]{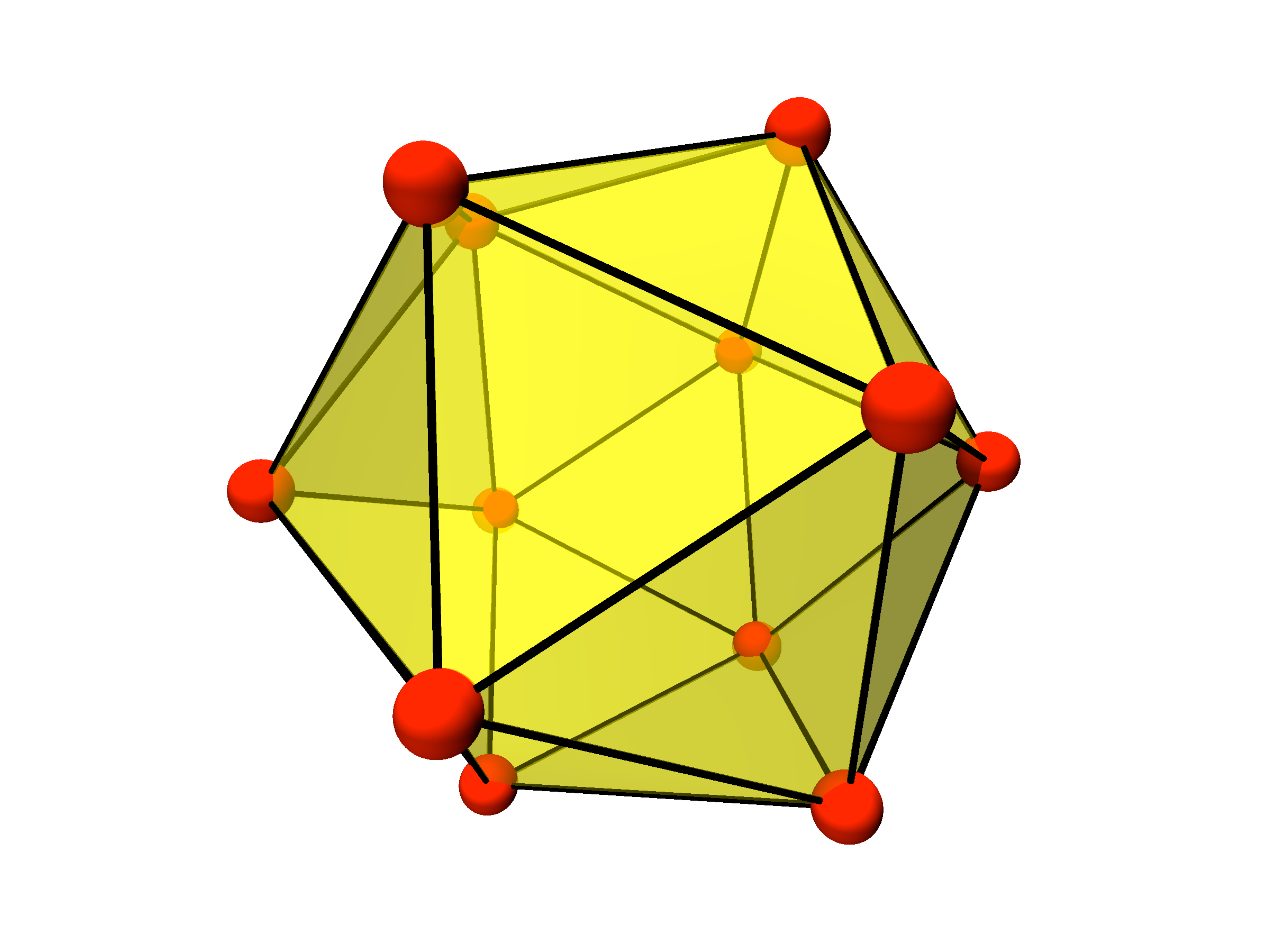}
\caption{Structure of the cuboctahedron (left) and the
  icosahedron (right). The bullets
  represent spin sites, the edges indicate interactions.} 
\label{tlmm-f-1}
\end{figure}

Before employing an approximation it is necessary to estimate
its accuracy by comparing to known exact results. For this
purpose we choose two highly frustrated model systems that have been
treated numerically exactly \cite{ScS:IRPC10,ScS:PRB09,ScS:P09,ScS:IRPC10}.
In both systems, the cuboctahedron and the icosahedron
(\figref{tlmm-f-1}), the spins are supposed to be mounted on the
vertices of the body. All spins interact antiferromagnetically
with their nearest neighbors, i.e. along the edges of the
body. The complete Hamiltonian of the spin system is given by
the Heisenberg and the Zeeman term, i. e. 
\begin{eqnarray}
\label{E-3-1}
\op{H}
&=&
-
2\;
\sum_{i<j}\;
{J}_{ij}
\op{\vec{s}}_i \cdot \op{\vec{s}}_j
+
g\, \mu_B\, B\,
\sum_{i}\;
\op{s}^z_i
\ .
\end{eqnarray}
${J}_{ij}$ is the exchange parameter between spins at sites $i$
and $j$. The antiferromagnetic case discussed in this article
corresponds to negative ${J}_{ij}$. For the sake of simplicity
it is assumed that all spins have the same spin quantum number
$s_1=s_2=\dots=s_N=s=3/2$ as well as the same $g$-factor, and
that ${J}_{ij}=J$ for nearest neighbors and zero otherwise.

\begin{table}[ht!]
\begin{center}
\begin{tabular}{|r|r|r|r|r|r|}
\hline
\hline
$M$ & $\text{dim}({\mathcal H}(M))$ & $R_1$ & $R_2$ & $R_3$ & $R_4$\\
\hline\hline
18 &        1 & exact & exact & exact & exact\\
17 &       12 & exact & exact & exact & exact\\
16 &       78 & exact & exact & exact & exact\\
15 &      364 & exact & exact & exact & exact\\
14 &     1353 & exact & exact & exact & exact\\
13 &     4224 & exact & exact & exact & exact\\
12 &    11440 & exact & exact & exact & exact\\
11 &    27456 & 1 & 5 & 20 & 100 \\
10 &    59268 & 1 & 5 & 20 & 100 \\
 9 &   116336 & 1 & 5 & 20 & 100 \\
 8 &   209352 & 1 & 5 & 20 & 100 \\
 7 &   347568 & 1 & 5 & 20 & 100 \\
 6 &   534964 & 1 & 5 & 20 & 100 \\
 5 &   766272 & 1 & 5 & 20 & 100 \\
 4 &  1024464 & 1 & 5 & 20 & 100 \\
 3 &  1281280 & 1 & 5 & 20 & 100 \\
 2 &  1501566 & 1 & 5 & 20 & 100 \\
 1 &  1650792 & 1 & 5 & 20 & 100 \\
 0 &  1703636 & 1 & 5 & 20 & 100 \\
\hline
\end{tabular}
\vspace*{5mm}
\end{center}
\caption{Employed number $R_i$ of random starting states for the
  cuboctahedron as well as the icosahedron with $s=3/2$: the
  columns provide the magnetic quantum number $M$, the
  dimensions of the subspaces ${\mathcal H}(M)$, and the
  $R_i$. ``exact" means that this subspace is included
  completely and exactly. The data for negative $M$ are given by
  the symmetry $M\leftrightarrow -M$.}\label{T-3-1} 
\end{table}

Since $\left[\op{H}, \op{{S}}^z\right] = 0$, this (simple)
symmetry is used for the finite-temperature Lanczos
calculations.  Table~\ref{T-3-1} shows, how the complete Hilbert
space is decomposed into subspaces ${\mathcal H}(M)$ with total
magnetic quantum number $M$. Besides the dimensions of those
subspaces the table also lists four scenarios $R_1$, $R_2$,
$R_3$, and $R_4$, that are used for the realization of the
FTLM. As mentioned earlier, small subspaces, here with $M\geq
12$, are treated exactly.

\begin{figure}[ht!]
\centering
\includegraphics*[clip,width=75mm]{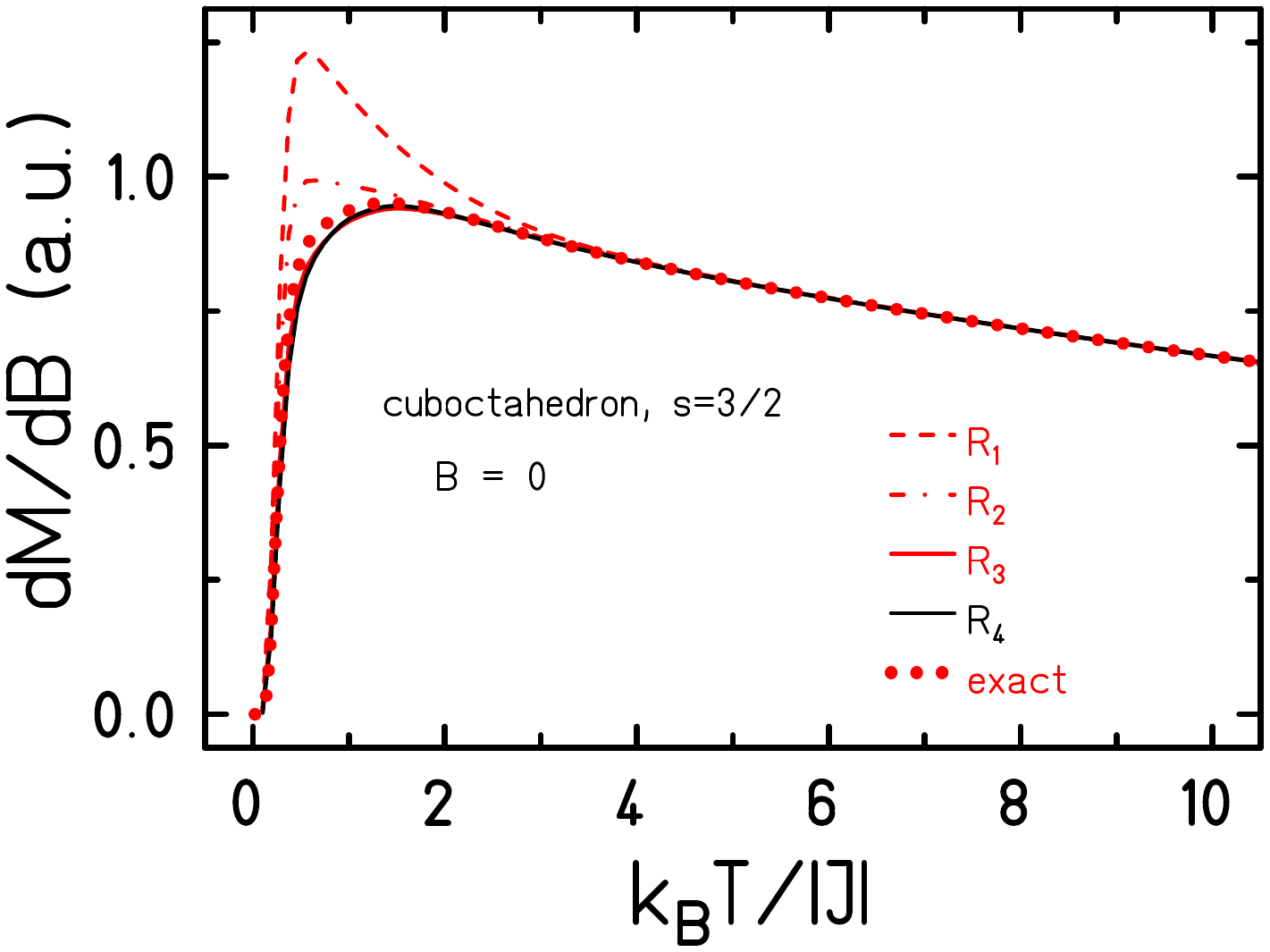}
\caption{Zero-field differential susceptibility of the
  cuboctahedron with $s=3/2$. The various curves depict the
  investigated scenarios $R_i$; $N_L=100$. The exact dependence is
  given by the dots.}
\label{tlmm-f-2}
\end{figure}

Figure \ref{tlmm-f-2} displays the zero-field differential
susceptibility of the cuboctahedron with $s=3/2$. One notices
that the approximate result, that anyway deviates from the exact
one only for $0.5 \leq k_B T/|J|\leq 3$, quickly approaches the
exact curve with increasing number $R$ of initial
states. Already for $R=20$ the approximation is practically
indistinguishable from the exact one; an increase to $R=100$
does not further improve this observable.

\begin{figure}[ht!]
\centering
\includegraphics*[clip,width=80mm]{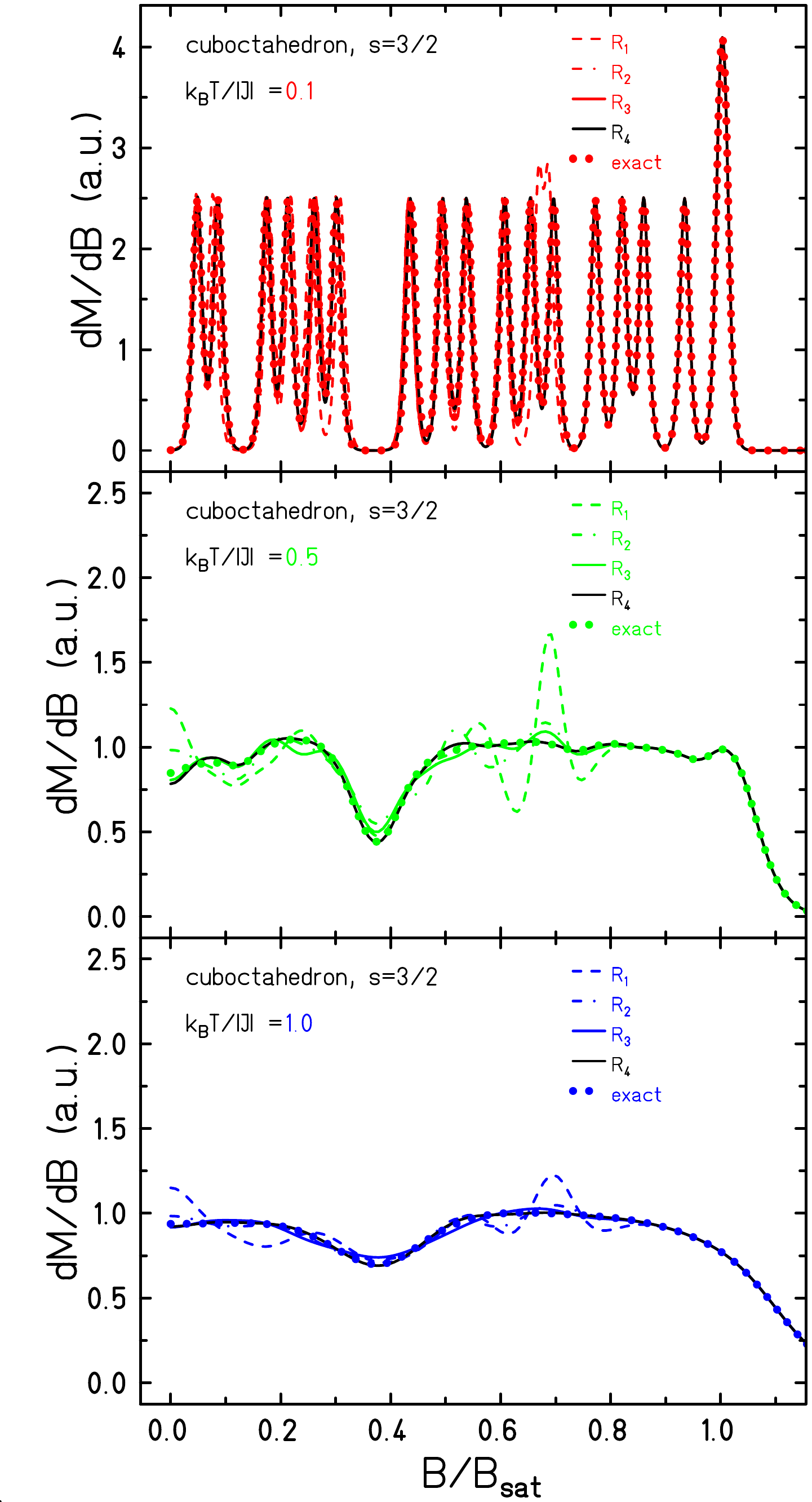}
\caption{Differential susceptibility of the
  cuboctahedron with $s=3/2$ for three temperatures. The various
  curves depict the 
  investigated scenarios $R_i$; $N_L=100$. The exact dependence is
  given by the dots.}
\label{tlmm-f-3}
\end{figure}

Figure \ref{tlmm-f-3} shows the same observable, but this time as a
function of the applied field for three low temperatures. Here
one clearly observes that $R_1=1$ leads to large deviations at
various fields. For $R_2=5$ the deviations are smaller but still
too big for a good approximation. The approximations for
$R_3=20$ and $R_4=100$ are again very good for the very low
temperature of $k_B T/|J|=0.1$ which is due to the fact that
low-lying levels which are dominant at this temperature are well
approximated with $N_L=100$ Lanczos steps. But for temperatures
of the order of the exchange interaction deviations can be
observed around the minimum at $B=B_{\text{sat}}/3$ for
$R_3=20$. This minimum is related to the magnetization plateau
with $M=M_{\text{sat}}/3$, see Refs.~
\cite{SSR:JMMM05,RLM:PRB08,Moe:JPCS09,Sch:DT10}.  It seems that
for smaller $R$ the higher-lying density of states is not quite
accurately reproduced in subspaces around $M=M_{\text{sat}}/3$.

\begin{figure}[ht!]
\centering
\includegraphics*[clip,width=75mm]{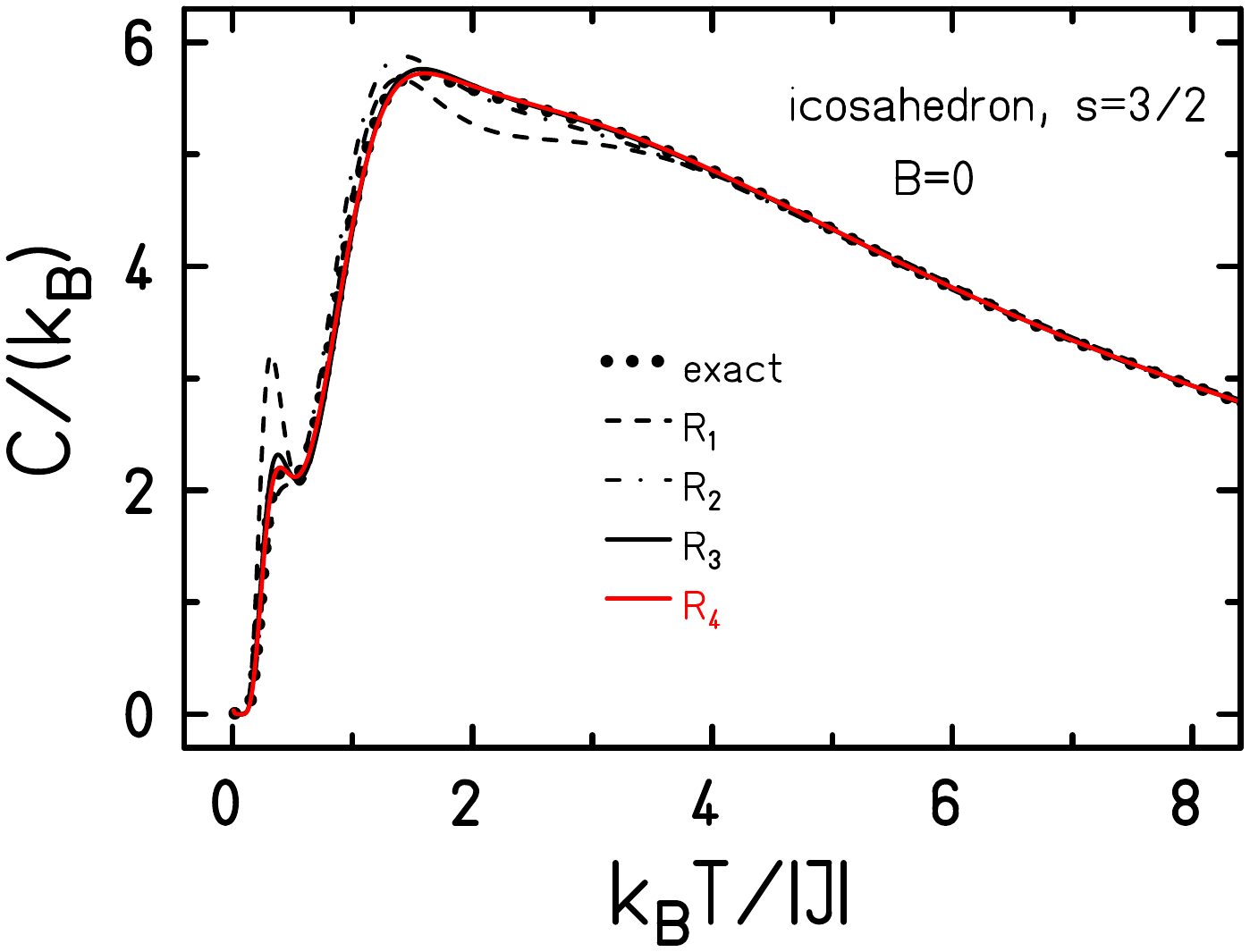}
\caption{Zero-field heat capacity of the
  icosahedron with $s=3/2$. The various curves depict the
  investigated scenarios $R_i$; $N_L=100$. The exact dependence is
  given by the dots.}
\label{tlmm-f-4}
\end{figure}

The magnetic properties of the icosahedron with spin $s=3/2$ are
discussed in Ref.~\cite{ScS:IRPC10}. Analogous to the
cuboctahedron its Heisenberg Hamiltonian can be diagonalized
numerically exactly with the help of point group
symmetries. Here we would like to compare the exact zero-field
heat capacity with the results of the finite-temperature Lanczos
method, again for the scenarios listed in Table~\ref{T-3-1}.
The reason to choose the heat capacity and not the
susceptibility is given by the fact that the heat capacity has
an unusual feature at $k_B T/|J|\approx0.5$ which can be
described as a small low-temperature Schottky peak. It stems
from a bunch of low-lying degenerate and nearly degenerate
energy levels \cite{ScS:IRPC10}. In addition the main maximum
has a rather unusual shape compared to other magnetic
molecules where the main maximum is sharper and as a function of
temperature drops off much more quickly towards the $1/T^2$
behavior at high temperatures. 

As one can see in \figref{tlmm-f-4} an approximation with just one
starting state ($R_1=1$) for each subspace is neither able to
reproduce the Schottky peak nor the main maximum. This is
already much better for $R_2=5$ and practically almost perfect
for $R_3=20$. We would like to emphasize once more that this
result is achieved with a very small number of states. As
Table~\ref{T-3-1} shows, the low-$M$ subspaces assume a size of
about 1.5 millions whereas the FTLM generates only $R\cdot N_L$
states in these subspaces, which for $R_3=20$ corresponds to
just 2,000 states. The calculations with $R_4=100$ practically
coincide with those for $R_3=20$; for the little Schottky peak
the accuracy is even further improved.

\section{Icosidodecahedron}
\label{sec-4}

\begin{figure}[ht!]
\centering
\includegraphics*[clip,width=45mm]{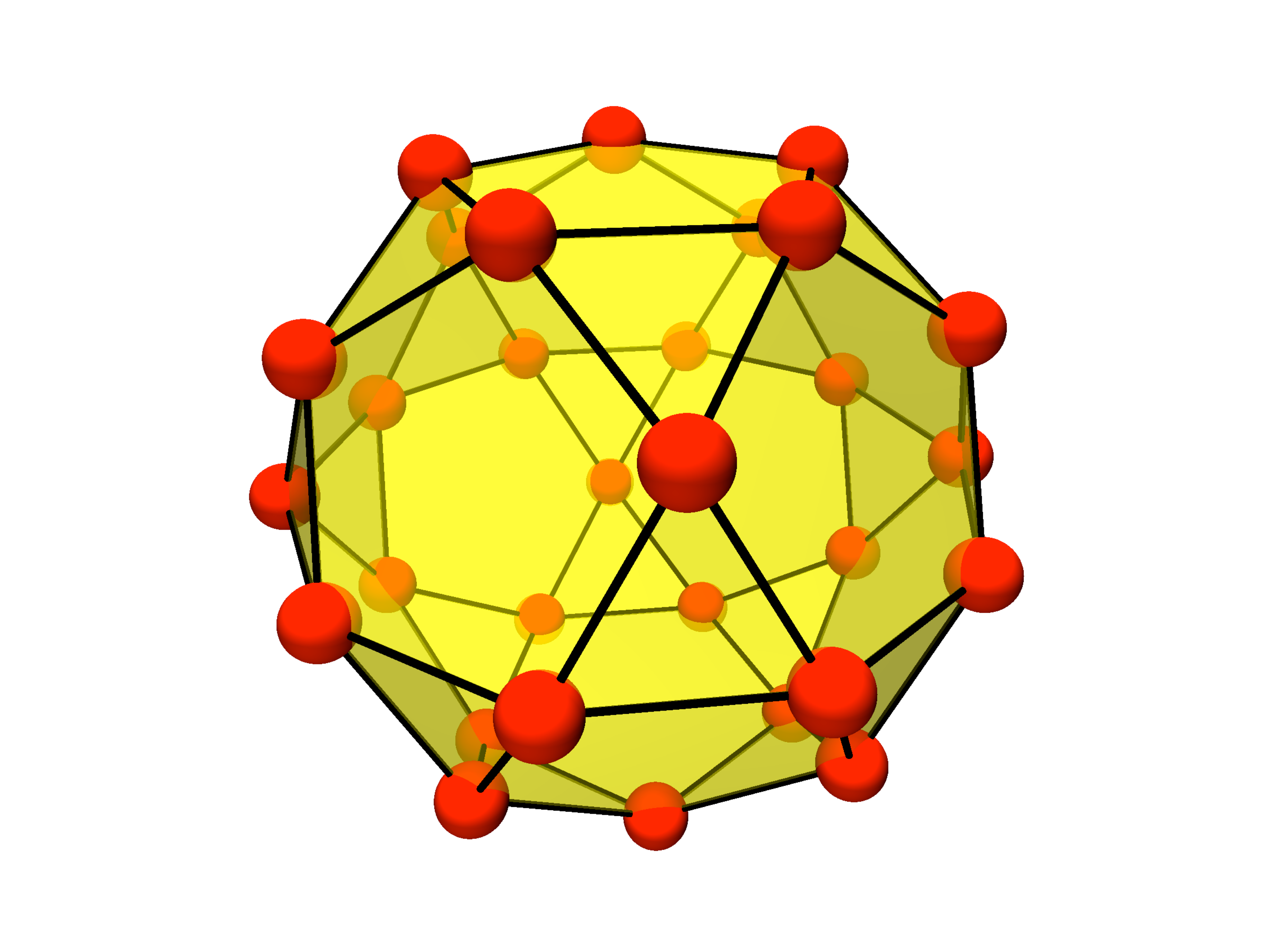}
\caption{Structure of the icosidodecahedron. The bullets
  represent the 30 spin sites, the edges indicate the 60
  exchange interactions. The structure is also termed
  Keplerate.}
\label{tlmm-f-5}
\end{figure}

The icosidodecahedron of antiferromagnetically coupled spins is
a very fascinating object, see \figref{tlmm-f-5} for the
structure. Chemically it is realized with spins $s=5/2$
(abbr. Fe$_{30}$ \cite{MSS:ACIE99}), $s=3/2$ (abbr. Cr$_{30}$
\cite{TMB:ACIE07}), and $s=1/2$ (abbr. V$_{30}$
\cite{MTS:AC05,BKH:CC05}). It belongs to the class of
geometrically frustrated kagom\'{e}-like spin systems. Due to
this close relation the icosidodecahedron once was termed ``The
kagom\'{e} on a sphere" \cite{RLM:PRB08}. These molecular
structures exhibit genuine properties of antiferromagnetic spin
systems built of corner sharing triangles as there are: many
singlet states below the first triplet state, a pronounced
magnetization plateau with $M/M_{\text{sat}}=1/3$, and a large
magnetization jump to saturation
\cite{SSR:EPJB01,SHS:PRL02,SSR:JMMM05,RLM:PRB08,Sch:DT10}.

\begin{table}[ht!]
\begin{center}
\begin{tabular}{|r|r|r|r|r|}
\hline
\hline
$M$ & $\text{dim}({\mathcal H}(M))$ & $R_1$ & $R_2$ & $R_3$\\
\hline\hline
15 &           1 & exact & exact & exact\\
14 &          30 & exact & exact & exact\\
13 &         435 & exact & exact & exact\\
12 &        4060 & exact & exact & exact\\
11 &       27405 & exact & exact & exact\\
10 &      142506 & exact & exact & exact\\
 9 &      593775 & 10 & 10 & 20 \\
 8 &     2035800 & 2 & 5 & 20 \\
 7 &     5852925 & 2 & 5 & 20 \\
 6 &    14307150 & 1 & 5 & 20 \\
 5 &    30045015 & 1 & 5 & 20 \\
 4 &    54627300 & 1 & 5 & 20 \\
 3 &    86493225 & 1 & 5 & 20 \\
 2 &   119759850 & 1 & 5 & 20 \\
 1 &   145422675 & 1 & 5 & 20 \\
 0 &   155117520 & 1 & 5 & 20 \\
\hline
\end{tabular}
\vspace*{5mm}
\end{center}
\caption{Employed number $R_i$ of random starting states for the
  icosidodecahedron with $s=1/2$: the
  columns provide the magnetic quantum number $M$, the
  dimensions of the subspaces ${\mathcal H}(M)$, and the
  $R_i$. ``exact" means that this subspace is included
  completely and exactly.}\label{T-4-1}  
\end{table}

Although these $(T=0)$ properties are accessible by means of
Lanczos diagonalization in the case of $s=1/2$
\cite{SSR:JMMM05,RLM:PRB08} and by means of DMRG calculations
for $s=3/2$ and $s=5/2$ \cite{ExS:PRB03}, the evaluation of the
thermal behavior, i.e. for $T>0$, seemed to be impossible due to
the prohibitive size of the Hilbert spaces. But at least for the
icosidodecahedron with $s=1/2$ the finite-temperature Lanczos
method could be able to deliver the temperature dependence of
the magnetic observables. Table~\ref{T-4-1} lists the parameters
used in our FTLM calculations. As can be deduced from the large
dimensions of the subspaces ${\mathcal H}(M)$ such calculations
are demanding. We employed the SGI Altix 4700 at the German 
Leibniz Supercomputing Center using openMP parallelization with
up to 510 cores as well as our local BULL/ScaleMP computer with
128 cores. To provide an estimate, a run in the subspace with
$M=0$ and $R_3=20$ together with $N_L=100$ needs about a full
day on 510 ITANIUM~II cores.

\begin{figure}[ht!]
\centering
\includegraphics*[clip,width=75mm]{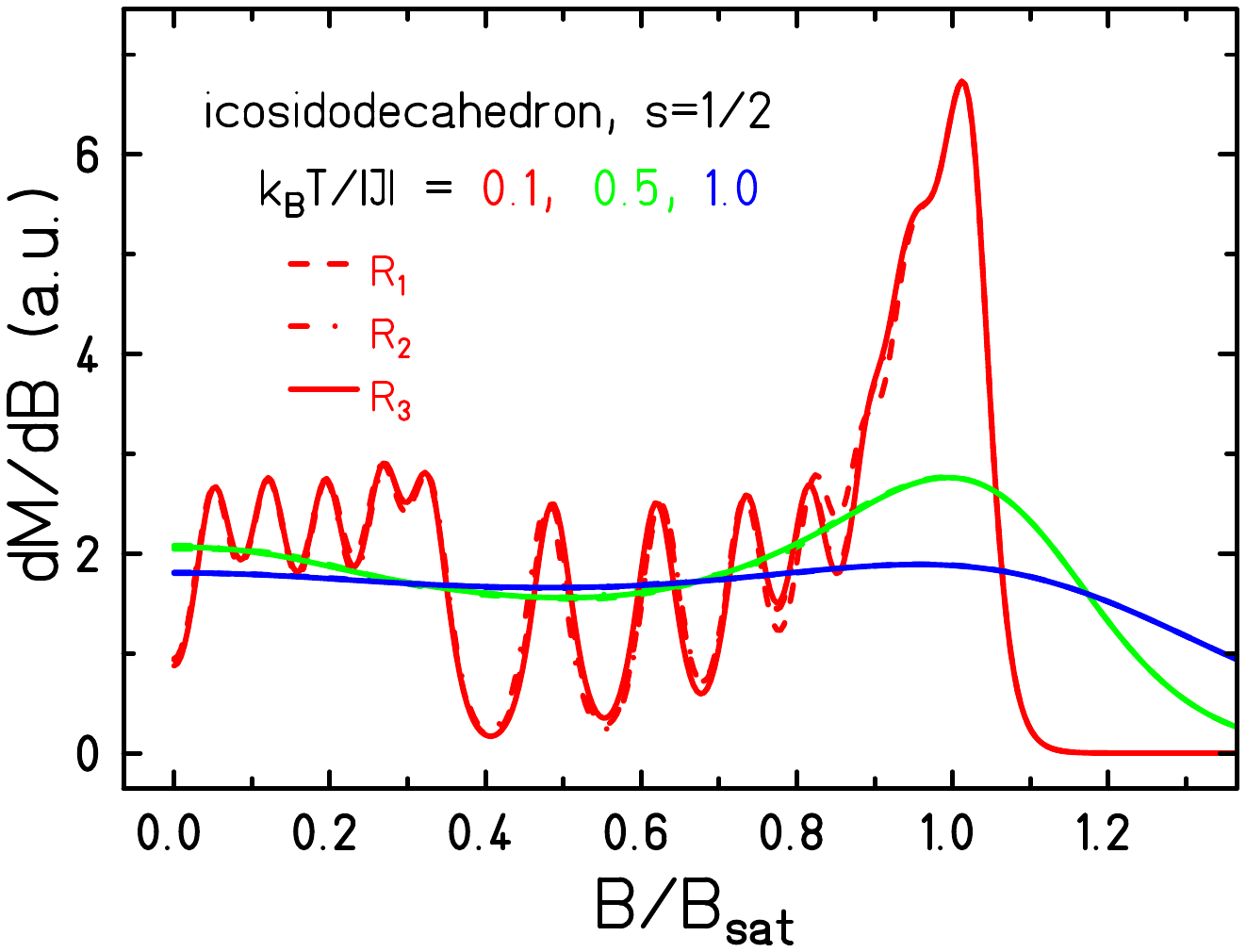}
\caption{Differential susceptibility of the icosidodecahedron with
  $s=1/2$ for three temperatures. The various curves depict the
  investigated scenarios $R_i$; $N_L=100$.}
\label{tlmm-f-6}
\end{figure}

Figure~\xref{tlmm-f-6} compares the results of FTLM calculations with
three different sets of random starting states, see
Table~\ref{T-4-1}. It is astonishing how little the observable
varies with $R_i$. This means that the finite-temperature
Lanczos method replaces the true spectrum very effectively by
pseudo energy eigenvalues so that gross properties are
efficiently reproduced.

\begin{figure}[ht!]
\centering
\includegraphics*[clip,width=75mm]{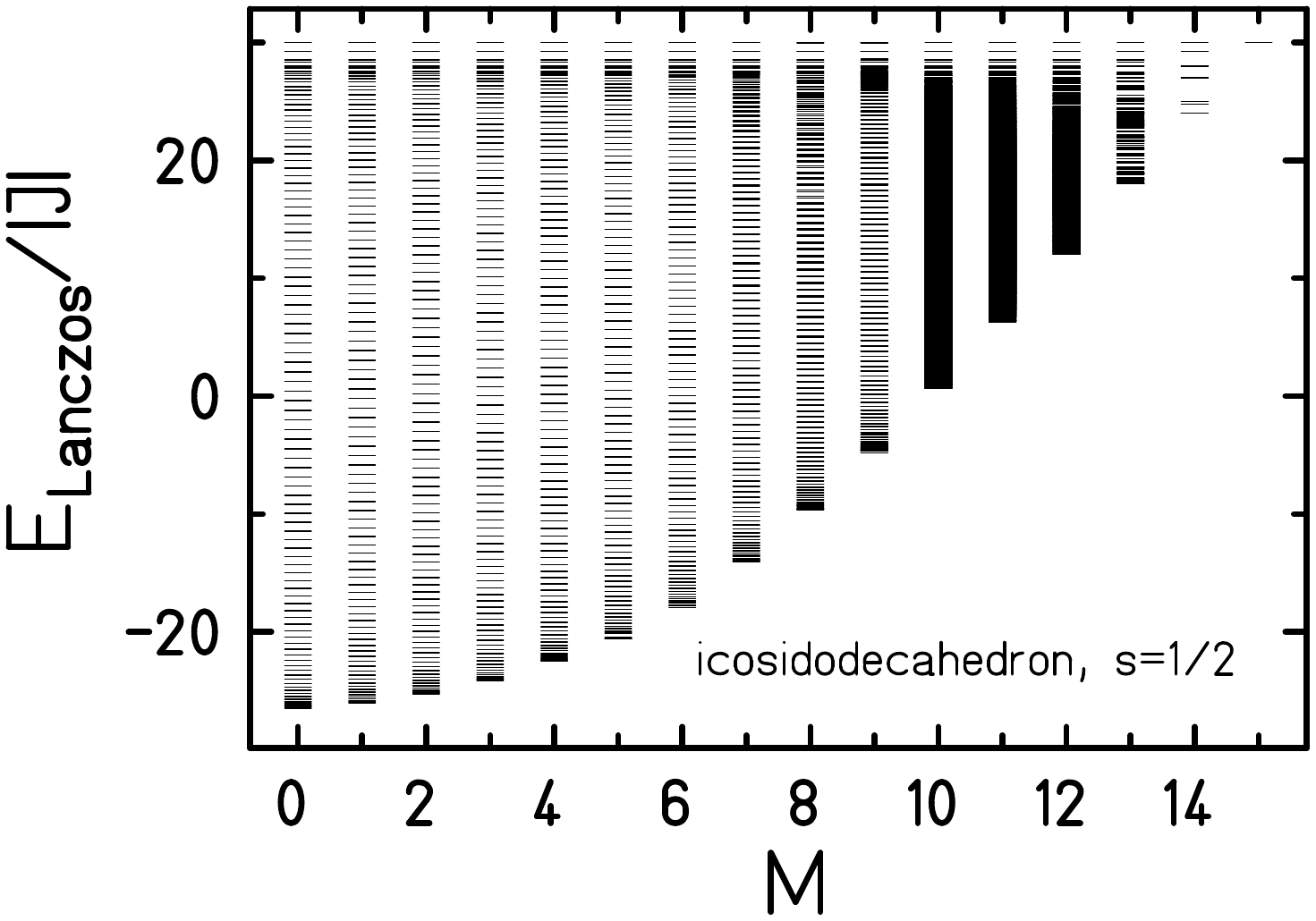}

\includegraphics*[clip,width=75mm]{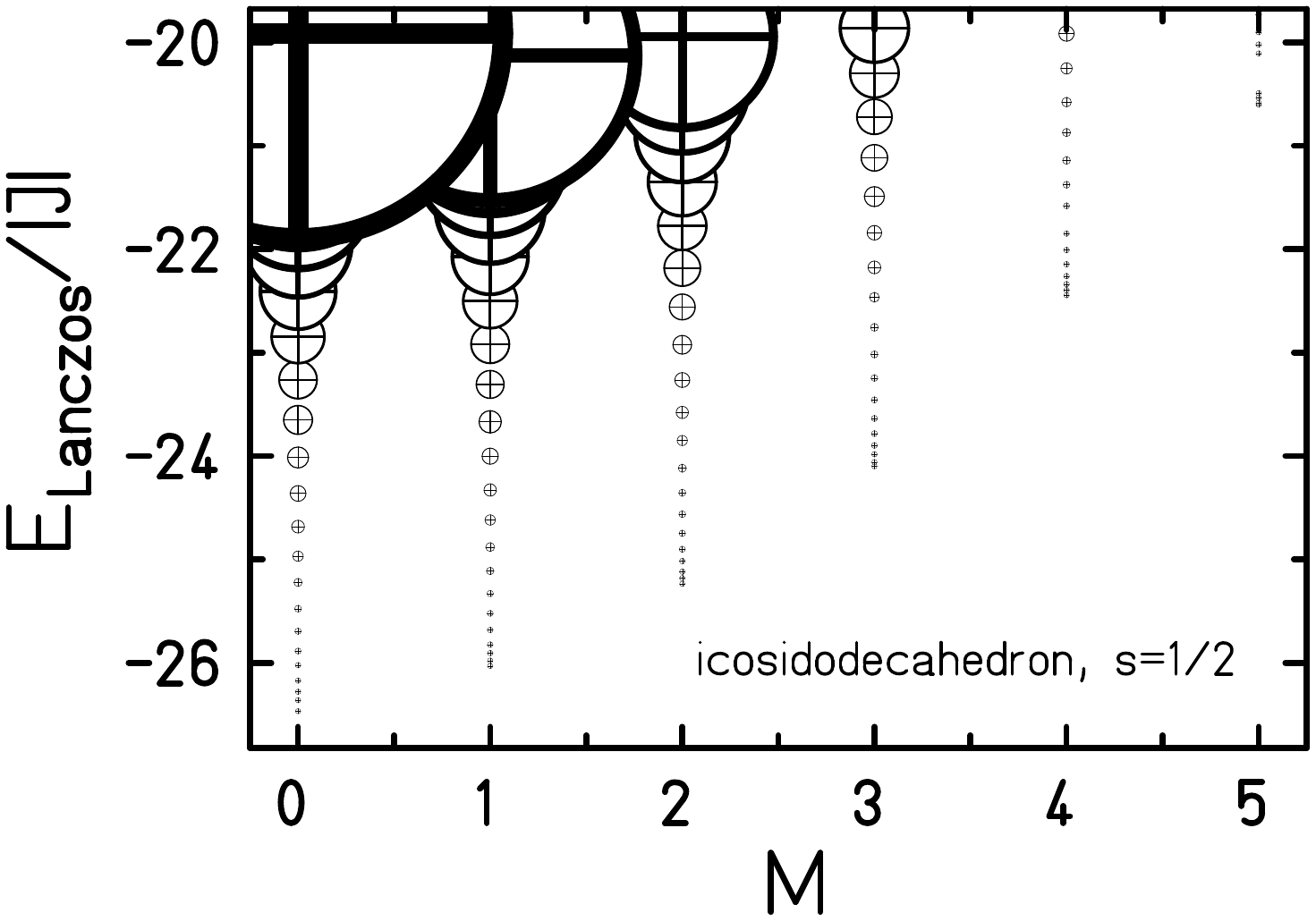}
\caption{Top: Pseudo energy eigenvalues of the icosidodecahedron with
  $s=1/2$ for $R_1=1$ and $N_L=100$. Bottom: Low-energy part of
  the same energy spectrum, but now the size of the symbols
  represents the weight of the corresponding state.}
\label{tlmm-f-7}
\end{figure}

It is important to note that the pseudo energy eigenvalues, see
top of \figref{tlmm-f-7}, have no spectroscopic meaning in
general. Very low-lying energy levels may nearly coincide with
the true ones due to the rapid convergence of the Lanczos method
for extremal eigenvalues. The vast majority of levels --
together with their weights(!) -- has to be understood as an
effective representation of the energy level density. To make
this point clearer the bottom of \figref{tlmm-f-7} displays the
low-energy part of the spectrum with symbols whose radii
represent the weights with which they have to be multiplied to
the Boltzmann factor in the partition function. In effect the
method has some similarities with the classical Wang-Landau
sampling \cite{WaL:PRL01,WaL:PRE01,ZST:PRL06}, where one also
constructs an approximate density of states consisting of
discretized energy intervals and their weights in order to later
evaluate thermal properties \cite{SBL:PRB07}.

\begin{figure}[ht!]
\centering
\includegraphics*[clip,width=80mm]{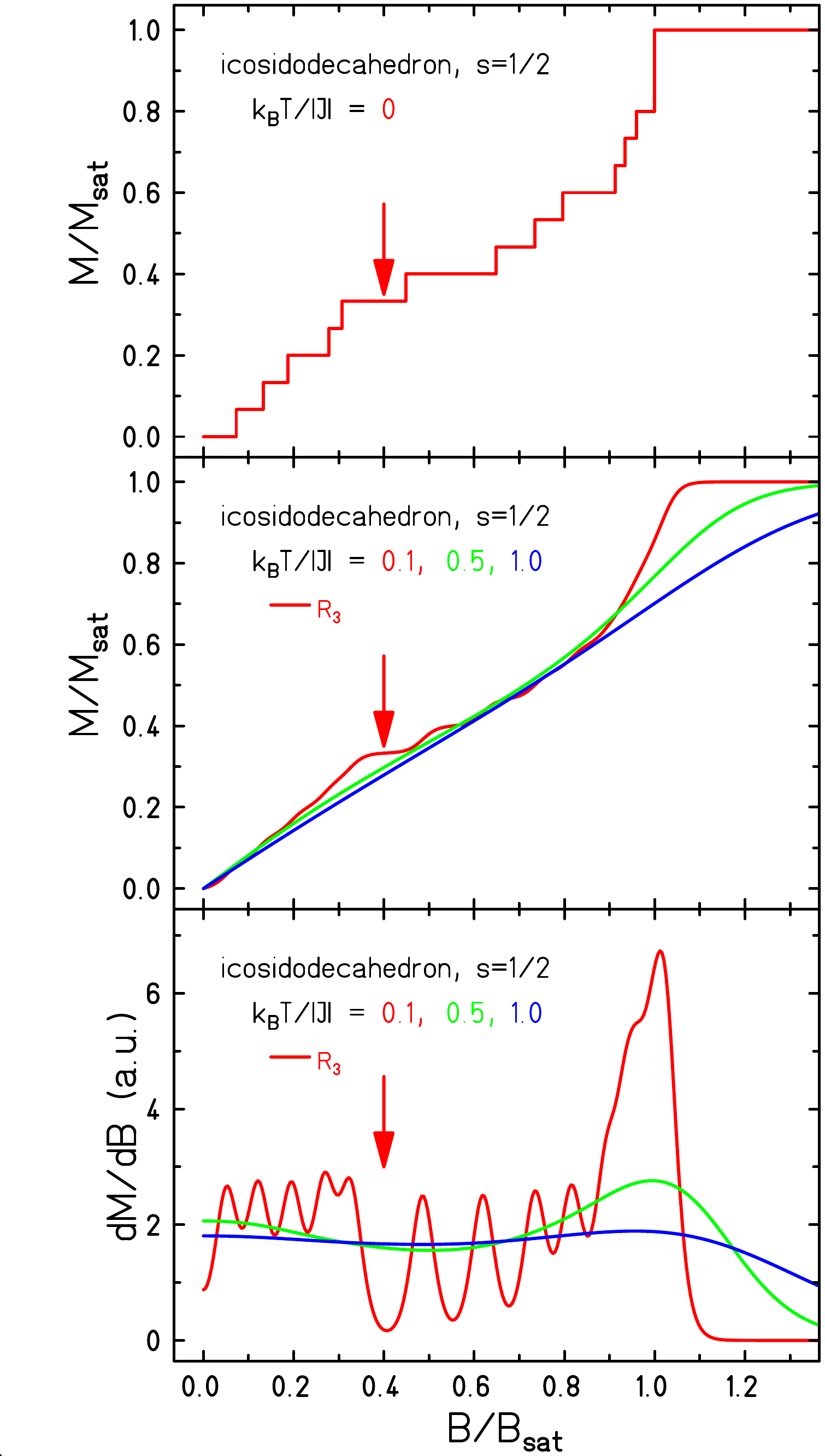}
\caption{Top and middle: Magnetization of the icosidodecahedron
  with $s=1/2$ for four temperatures and $R_3=20$; $N_L=100$. The
  plateau with $M/M_{\text{sat}}=1/3$ is indicated by an
  arrow. The other steps of the magnetization curve are due to
  the finite size of the spin system.
  Bottom: corresponding differential susceptibility;
  here the plateau expresses itself as a dip.}
\label{tlmm-f-8}
\end{figure}

The successful determination of the temperature dependence
enables us to discuss several thermal properties of the
icosidodecahedron with $s=1/2$. A key question is the width and
thermal evolution of the magnetization plateau with
$M/M_{\text{sat}}=1/3$. This plateau expresses itself in the
differential susceptibility as a dip. Classical calculations for
the $s=5/2$ case yielded a dip that is much narrower than the
experimental findings \cite{SNS:PRL05}. It was not evident how
this feature would behave in a quantum
calculation. Figure~\xref{tlmm-f-8} shows both the magnetization (top
and middle) as well as the differential susceptibility
(bottom). The plateau with $M/M_{\text{sat}}=1/3$ is indicated
by an arrow. Interestingly, the plateau as well as the dip
disappear quickly with rising temperature. Already for $k_B
T/|J|=0.5$ they are hardly visible, and the position of the now
much broader dip is shifted to higher fields. It is not yet
obvious -- and thus will be a matter of future research -- how
this trend transfers to quantum icosidodecahedra with $s=5/2$ or
$s=3/2$ and whether it would be sufficient to explain the
experimental findings \cite{SPK:PRB08}.
 
\begin{figure}[ht!]
\centering
\includegraphics*[clip,width=75mm]{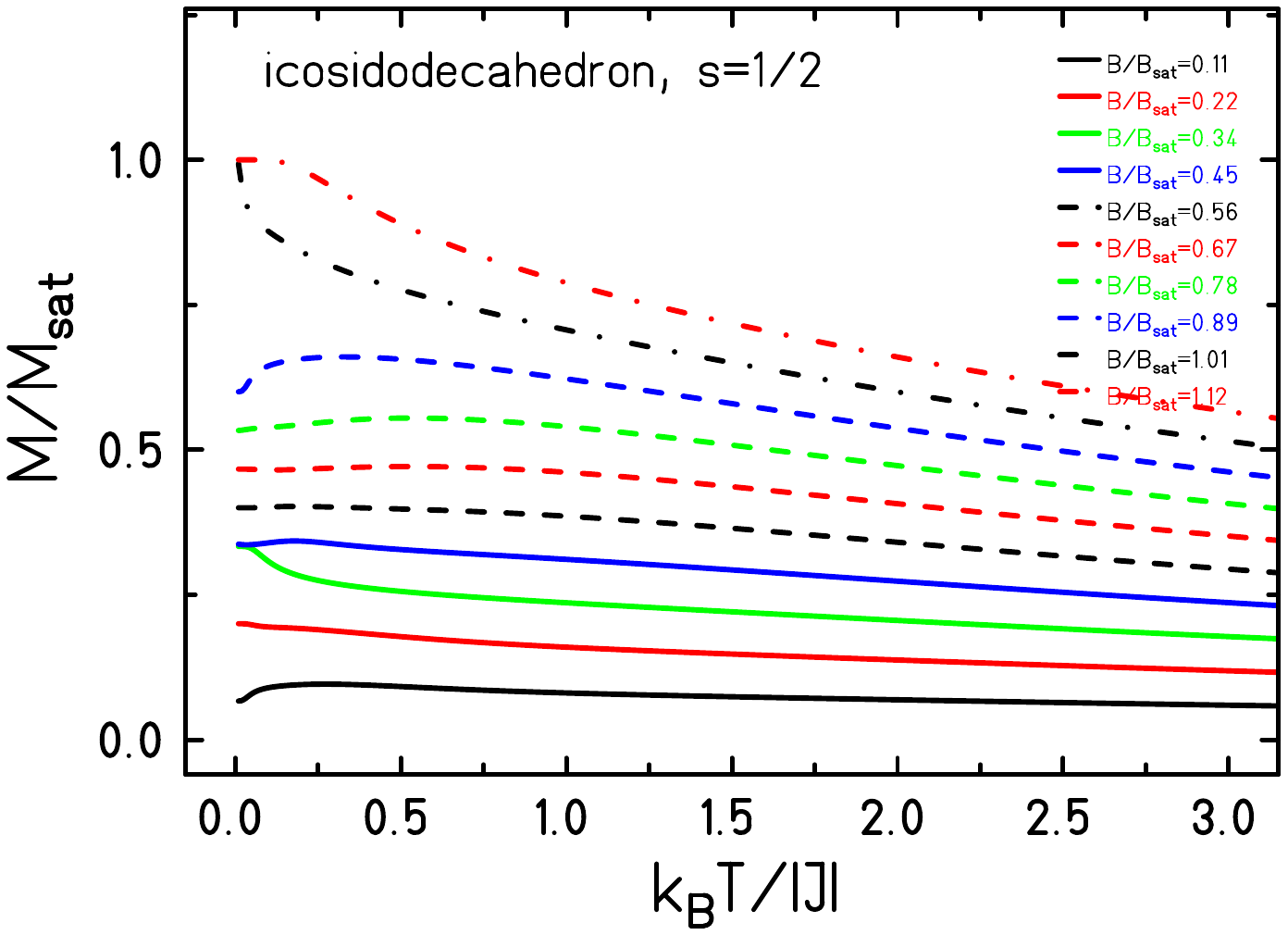}
\caption{Magnetization of the icosidodecahedron with $s=1/2$ for
  various magnetic field strengths and $R_3=20$; $N_L=100$.}
\label{tlmm-f-9}
\end{figure}

Another feature, that was observed for the icosidodecahedron with
$s=5/2$, is the near constancy of the magnetization as a function
of temperature for some magnetic fields
\cite{SLM:EPL01}. Figure~\xref{tlmm-f-9} displays the theoretical
magnetizations for low temperatures and several magnetic field
strengths, but now of course for $s=1/2$. There are field
ranges, e.g. $B/B_{\text{sat}}\approx 0.4, \dots, 0.8$, where
the magnetization varies indeed very little with
temperature. This is also seen in the middle part of
\figref{tlmm-f-8}, where the magnetization curves for three
temperatures virtually fall on top of each other in the
respective field interval. At the moment it is a speculation how
this behavior would change for larger spins such as $s=5/2$. One
could conjecture that the field ranges of the plateau as well as
of the rise to saturation might shrink relatively and thus lead
to thermally stable magnetization in broader field intervals.

\begin{figure}[ht!]
\centering
\includegraphics*[clip,width=75mm]{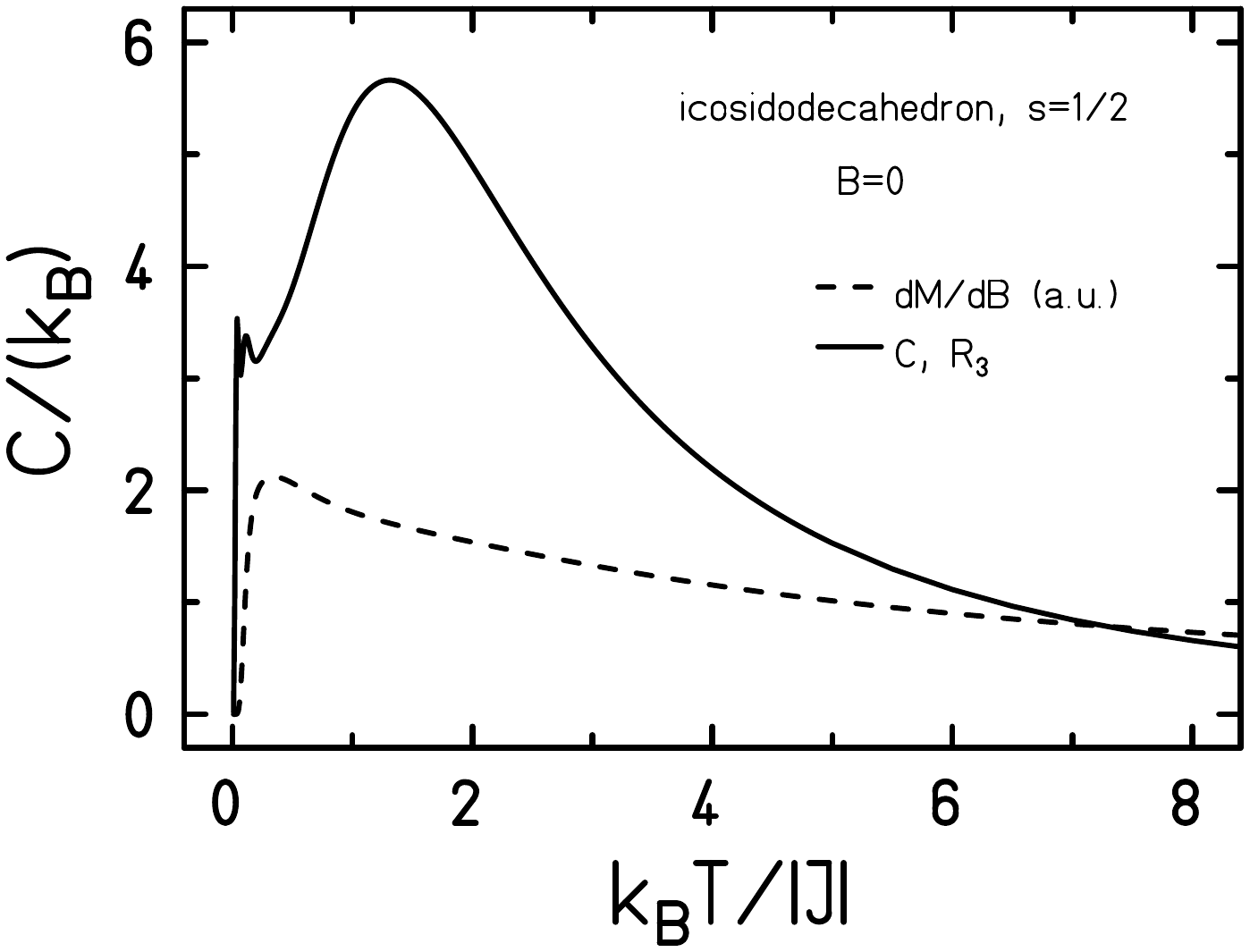}
\caption{Zero-field heat capacity of the icosidodecahedron with
  $s=1/2$ and $R_3=20$; $N_L=100$ (solid curve). For comparison
  the zero-field differential susceptibility is given by a
  dashed curve.}
\label{tlmm-f-10}
\end{figure}

Finally, we would like to discuss the heat capacity of the
icosidodecahedron with $s=1/2$. Figure~\xref{tlmm-f-10} shows the
zero-field heat capacity that is obtained for $R_3=20$ and
$N_L=100$. Since the system possesses many singlets below the
first triplet one expects low-temperature features in the
specific heat curve, that are indeed clearly visible (solid
curve). They are absent in the zero-field differential
susceptibility (dashed curve) that is provided for
comparison. In addition the main maximum of the specific heat is
at higher temperatures than that of the susceptibility which
points at a higher-lying density of states that shows up in the
heat capacity but has not much impact on the susceptibility.

\section{Summary and Outlook}
\label{sec-5}

The finite-temperature Lanczos method enabled us to evaluate the
thermal properties of the antiferromagnetic spin
icosidodecahedron with $s=1/2$. The magnetic susceptibility as
well as the heat capacity could be determined. A major result is
that the magnetization plateau at $M/M_{\text{sat}}=1/3$ is
thermally rather unstable, i.e. it disappears above temperatures
of $k_B T/|J|\approx 0.5$.

An important open question is how our findings change if the
spin is increased, e.g. to $s=5/2$ for the iron based
icosidodecahedron. Intuitively one would guess that quantum
features are decreased for the more classical spin. For the
smaller but similar cuboctahedron it could be shown that the
number of singlets below the first triplet state decreases with
increasing spin quantum number \cite{ScS:P09}. This would have
an impact on the low-temperature features of heat
capacity. Further investigations are necessary to clarify such
questions which are of general nature for all kagom\'{e}-like
spin systems.

\section*{Acknowledgment}

This work was supported by the German Science Foundation (DFG)
through the research group 945. Computing time at the Leibniz
Computing Center in Garching is also gratefully
acknowledged. Last but not least we like to thank the State of
North Rhine-Westphalia and the DFG for financing our local SMP
supercomputer as well as the companies BULL and ScaleMP for
their support.

\end{document}